\documentclass{emulateapj}

\begin{document}

\title{GALEX UV observations of the interacting galaxy NGC 4438 in the Virgo cluster.}
\author{
A. Boselli\altaffilmark{1}, S. Boissier\altaffilmark{2}, L. Cortese\altaffilmark{1},  
A. Gil de Paz\altaffilmark{2}, V. Buat\altaffilmark{1}, J. Iglesias-Paramo\altaffilmark{1},
B. F. Madore\altaffilmark{2,8},
T. Barlow\altaffilmark{3}, L. Bianchi\altaffilmark{4},
Y.-I. Byun\altaffilmark{5}, J. Donas\altaffilmark{1},
K. Forster\altaffilmark{3}, P. G. Friedman\altaffilmark{3},
T. M. Heckman\altaffilmark{6}, 
P. Jelinsky\altaffilmark{7},
Y.-W. Lee\altaffilmark{5}, 
R. Malina\altaffilmark{1},
D. C. Martin\altaffilmark{3}, B. Milliard\altaffilmark{1},
P. Morrissey\altaffilmark{3}, S. Neff\altaffilmark{9},
R. M. Rich\altaffilmark{10}, D. Schiminovich\altaffilmark{3},
M. Seibert\altaffilmark{3},
O. Siegmund\altaffilmark{7}, 
T. Small\altaffilmark{3},
A. S. Szalay\altaffilmark{6}, 
B. Welsh\altaffilmark{7}, T. K. Wyder\altaffilmark{3}
}
\altaffiltext{1}{Laboratoire d'Astrophysique de Marseille, BP8, Traverse du Siphon, F-13376 Marseille, France}
\altaffiltext{2}{Observatories of the Carnegie Institution of Washington,
813 Santa Barbara St., Pasadena, CA 91101}
\altaffiltext{3}{California Institute of Technology, MC 405-47, 1200 East
California Boulevard, Pasadena, CA 91125}
\altaffiltext{4}{Center for Astrophysical Sciences, The Johns Hopkins
University, 3400 N. Charles St., Baltimore, MD 21218}
\altaffiltext{5}{Center for Space Astrophysics, Yonsei University, Seoul
120-749, Korea}
\altaffiltext{6}{Department of Physics and Astronomy, The Johns Hopkins
University, Homewood Campus, Baltimore, MD 21218}
\altaffiltext{7}{Space Sciences Laboratory, University of California at
Berkeley, 601 Campbell Hall, Berkeley, CA 94720}
\altaffiltext{8}{NASA/IPAC Extragalactic Database, California Institute
of Technology, Mail Code 100-22, 770 S. Wilson Ave., Pasadena, CA 91125}
\altaffiltext{9}{Laboratory for Astronomy and Solar Physics, NASA Goddard
Space Flight Center, Greenbelt, MD 20771}
\altaffiltext{10}{Department of Physics and Astronomy, University of
California, Los Angeles, CA 90095}

\begin{abstract}
We present GALEX NUV ($2310\rm \AA$) and FUV ($1530\rm \AA$) images of the interacting galaxy NGC 4438 (Arp 120) in the center of the 
Virgo cluster. These images show an extended (20 kpc) tidal tail at the north-west edge of the galaxy 
previously undetected at other wavelengths, at 15-25 kpc from its nucleus. 
Except in the nucleus, the UV morphology of NGC 4438 is totally different from 
the H$\alpha$+[NII] one, more similar to the X-ray emission,
confirming its gas cooling origin. We study the star formation history of NGC 4438 
combining spectro-photometric data in the UV-visible-near-IR wavelength range with population synthesis and galaxy evolution 
models. The data are consistent with a recent ($\sim$ 10 Myr), instantaneous burst of star formation in the newly discovered 
UV north-western tail which is significantly younger than the age of the tidal interaction with NGC 4435, 
dated by dynamical models at $\sim$ 100 Myr ago. Recent star formation events are also present at the edge of the northern 
arm and in the southern tail, while totally lacking in the other regions, which are dominated by the old stellar population 
perturbed during the dynamical interaction with NGC 4435. The contribution of this recent starburst to the total galaxy stellar mass
is lower than 0.1\%, an extremely low value for such a violent interaction. High-velocity, off-center tidal encounters such as that observed in
Arp 120 are thus not sufficient to significantly increase the star formation activity of cluster galaxies.

\end{abstract}


\section{Introduction}

\setcounter{footnote}{0}

NGC 4438 (Arp 120) is the clearest example of an ongoing tidal interaction in a nearby cluster of galaxies. Apparently located close to
the Virgo cluster center ($\sim$ 300 kpc from M87), NGC 4438 is a bulge-dominated late-type spiral showing long tidal
tails (30 kpc) thought to be induced by a recent dynamical interaction with the nearby SB0 galaxy NGC 4435.
Multifrequency observations covering the electromagnetic spectrum from X-rays (Kotanyi et al$.$ 1983; Machacek et al$.$ 2004)
to radio continuum (Hummel \& Saikia 1991), including both 
spectro-photometric and kinematical (Kenney et al$.$ 1995) data, have been carried out in the past to study the nature of 
this peculiar system.
These observations have shown that the violent interaction between the two galaxies perturbed the atomic 
(Cayatte et al$.$ 1990) and molecular (Combes et al$.$ 1988) 
gas distribution, causing both gas infall toward the center which might have induced nuclear activity (Kenney et al$.$ 1995;
Kenney \& Yale 2002; Machacek et al$.$ 2004), and gas removal in the external parts displacing part of the gas in the ridge in between the
two galaxies (Combes et al$.$ 1988).\\
Both multifrequency observational data (Kenney et al$.$ 1995; Machacek et al$.$ 2004) and model 
predictions (Combes et al$.$ 1988; Vollmer et al$.$ 2005) favor a recent 
($\sim$ 100 Myr) high-velocity, off-center collision between NGC 4435 and NGC 4438.\\
Except for mild nuclear activity, it is still unclear whether the dynamical interaction between the two galaxies 
induced extra-nuclear star formation events: the low H$\alpha$/[NII] ratio and the similar X-ray and H$\alpha$ morphology
of NGC 4438 indicate that the H$\alpha$ emission 
is in this case not due to the ionizing radiation but is probably due to gas cooling phenomena (Machacek et al$.$ 2004).\\
The UV emission (at $\sim$ 2000 \AA) is dominated by young stars of intermediate masses ($2<M<5M_{\odot}$, e.g. Boselli et al$.$ 2001)
and provides us with an alternative star formation tracer. As part of the Nearby Galaxy Survey (NGS), we have observed 
the central 12 deg$^2$ of the Virgo cluster using the \emph{Galaxy Evolution Explorer} (GALEX). In this paper we present the 
UV observations of the interacting galaxies NGC 4435 and NGC 4438 \footnote{A 
distance of 17 Mpc for Virgo is adopted}.

\section{Data}

The GALEX data used in this work include far-ultraviolet (FUV; $\rm \lambda_{eff}=1530\AA, \Delta \lambda=400\AA$) and near-ultraviolet 
(NUV; $\rm \lambda_{eff}=2310\AA, \Delta \lambda=1000\AA$) images with a circular field of view of radius $\sim$ 0.6 degrees.
The spatial resolution is $\sim$5 arcsec. Details of the GALEX instrument and data characteristics can be found in 
Martin et al$.$ (2005) and Morrissey et al$.$ (2005).
The data (IR0.2 release) consist of 2 independent GALEX pointings centered at R.A.(J2000)=12h29m01.2s, Dec(J2000)=+13$^{\circ}$10'29.6" (819 sec) and R.A.(J2000)=12h25m25.2s, 
Dec(J2000)=13$^{\circ}$10'29.6" (1511 sec), for a total of 2330 sec of integration time.\\
To study the star formation history of NGC 4438, the UV data have been combined with visible and near-IR images of the galaxy taken from the 
GOLDmine database (http:\slash \slash goldmine.mib.infn.it; Gavazzi et al$.$ 2003), from the SDSS Data release 3 (Abazajian et al$.$ 2004) 
from the 2MASS survey (Jarrett et al$.$ 2003) and from the CFHT and SUBARU archives. 
These are H$\alpha$+[NII] (Boselli \& Gavazzi 2002), B (Boselli et al$.$ 2003a), K' (Boselli et al$.$ 1997), 
$u,g,r,i,z$ SDSS, R CFHT and SUBARU and H 2MASS images.
For the main body of the galaxy (region 4 in Fig. 1, see next sect.) we added the integrated spectrum (3500-7000 \AA~; Gavazzi et al$.$ 2004).
The current calibration errors of the NUV and FUV magnitudes are on the order of $\sim 10$\% (Morrissey et al$.$ 2004), comparable to
that at other frequencies. 

\section{The UV emission and the star formation history of NGC 4438}

Figure 1 shows the UV image of NGC 4438, obtained by combining together the NUV and FUV frames in order to 
increase the S/N. The UV emission of the galaxy is mostly due to compact, bright regions in the
central part of the galaxy (marked as region 4 in Fig. 1), in the northern tidal tail (region 2) and in the section 
of the southern tail closest to the main body of the galaxy (region 5). The UV emission is mostly diffuse
in the extended western part of the northern tail (region 3) and at the edge of the southern tidal tail (region 6).
Figure 1 shows the presence of extended and patchy emission to the north-west of the galaxy ($\sim$ 15-25 kpc from the nucleus,
marked as region 1). This feature, previously undetected in other visible and/or near-IR bands, is similar to a 
tidal tail $\sim$ 20 kpc long and $\sim$ 2 kpc wide. A similar smaller region ($\sim$ 2 kpc) 
at $\sim$  25 kpc from the nucleus is designated region 7. The average NUV surface brightness 
of these features is $\sim$ 28.5 ABmag $\rm arcsec^{-2}$, while they
are undetected in the SUBARU R band (360 sec) image down to a surface brightness limit of $\sim$ 27.8 mag $\rm arcsec^{-2}$.

The RGB image of the galaxy obtained by combining the FUV, NUV and B frames, given in Plate 1, shows the color of the different
regions: while the edge of both the northern and the southern tidal tails (region 3 and 6) are red (thus dominated by relatively old stars),
regions 2 and 5 as well as the newly discovered regions 1 and 7, have blue 
colors and seem therefore to be dominated by a younger population.
The H$\alpha$+[NII] emission map, given in Fig. 2 as a contour plot superposed on the NUV image of NGC 4438, 
shows a lack of massive, ionizing young O-B stars\footnote{The H$\alpha$+[NII] 
emission observed in region 5 has a different morphology than the UV one: this evidence
confirms the conclusions of Machacek et al$.$ (2004) that the H$\alpha$+[NII] emission is not due to the ionizing radiation but is
probably associated to the cooling gas} 
(Kennicutt 1998) down to a surface brightness limit of $\sim$ 5 10$^{-17}$ erg s$^{-1}$ cm$^{-2}$ arcsec$^{-2}$. \\
Extraplanar diffuse regions with an excess of UV over H$\alpha$ flux ratio (as that observed at 11 kpc from
the disk of M82) are often interpreted as due to the UV radiation produced by the central starburst and locally scattered
by diffuse dust (Hoopes et al$.$ 2005). It is unlikely that scattered light is responsible for the UV emission in regions 2 and 5 
since it comes from disk HII regions. The steep slope of the UV spectrum ($\beta$=-2.32 and -2.05, as defined by Kong et al$.$ 2004
respectively for regions 1 and 7) is typical of a recent unreddened starburst (Calzetti 2001) and 
is unexpected in a scattering scenario since the dust albedo is greater in
the NUV than in the FUV (Draine 2003). Furthermore the lack of a powerful central starburst (as in M82) and
the large distance of these relatively patchy regions from the nucleus seem to exclude the scattering scenario.\\
These data suggest that regions 1 and 7 (and to a lesser degree regions 2 and 5) are post
starbursts, induced by the violent interaction with NGC 4435, but that they lasted for a relatively short time, 
since they are not producing young, massive O-B stars any more. This is 
probably because the atomic and molecular gases, needed to feed star formation, have been removed during the interaction 
(Combes et al$.$ 1988; Vollmer et al$.$ 2005) \footnote {The upper limit of the HI surface density for these regions
is $\sim$ 1 M$\odot$ pc$^{-2}$ (Cayatte et al$.$ 1990)} .\\

In order to age-date the starburst and reconstruct the star formation history of the galaxy, 
we have determined the spectral energy distribution (SED) of these seven regions (see Plate 1) and then fitted them with a model
of galaxy evolution. To this end we make the assumption that dust attenuating the SED is present only in region 4, where 
we correct the UV to near-IR data using the far-IR to UV flux ratio as done in Boselli et al$.$ (2003a). This restricted application is reasonable
since no dust emission has been observed in the tidal tails with ISOCAM (Boselli et al$.$ 2003b); furthermore, in regions 1 and 7, 
dust is unexpected since it has not yet been produced by the young stellar population, as confirmed by the steep $\beta$ parameter.\\
We assume that NGC 4438 was a normal, late-type object before interacting with NGC 4435. 
The models of chemo-spectrophotometric
evolution of normal, spiral galaxies of Boissier \& Prantzos (2000), updated
with an empirically-determined star formation law (Boissier et al$.$
2003) are used to reconstruct the SED of the galaxy stellar population before the interaction.   
The two parameters of the model (spin $\lambda$ and rotational velocity
$V_C$) are constrained by the total H-band luminosity, the velocity
rotation and by the SED of the main body of the galaxy (region 4),
composed by an old population with no significant contribution from the
recent starburst (leading to $\lambda$=0.01 and $V_C$=290 km s
$^{-1}$).
We then assume that the underlying stellar population of each 
region, if present, is the one given by the model and removed
from the main body of the galaxy by the tidal interaction, while the younger population
is produced by the induced starburst. 
For each region, we combine the evolved stellar population 
with an instantaneous burst of star formation obtained using Starburst 99
(Leitherer et al$.$ 1999) for a solar metallicity and a Salpeter IMF 
between 1 and 100 $M_{\odot}$. For each age and intensity of the burst, 
we determine the best combination of evolved population+ burst
by fitting the FUV to K band SED and rejecting solutions 
in disagreement with the upper limits. We then adopt the age corresponding to the lowest reduced $\chi^2$ \footnote{All ages with
$\chi^2$ $<$ 1 are acceptable solutions.Given the small number of photometric points available for regions 1 and 7 (2 GALEX bands),
the fitted solution for a combination of a burst and an old population (two parameters) can
be almost perfect (resulting in very low $\chi^2$, $\leq$ 10$^{-2}$), as long as  
the obtained fit is in agreement with the limits at other wavelengths. 
Whenever the fit produces a SED not satisfying a detection limit, this
solution is rejected.}.  
This exercise gives a consistent result: the strong UV emission of regions 1 and 7 
is due to a coeval starburst $\sim$ 6-20 Myr old (see Panel 1). The age and the duration of the starburst
are strongly constrained both by the lack of H$\alpha$ emission and by the blue UV slope 
of the spectrum (lower limit to the age) and by the lack of an old stellar population (upper limit to the duration).
Regions 2 and 5 are well fitted by an older starburst ($\geq$ 100 Myr)
which ended $\sim$ 10 Myr ago as indicated by the lack of any H$\alpha$ emission and a redder
UV slope ($\beta$=-0.33 and -0.67 in regions 2 and 5 respectively).
While the fraction of stars produced by this burst is
dominant in regions 1 and 7, the sum of the stars produced by the burst in all regions (including the inner part) 
contributes to the total galaxy stellar mass by less than 0.1 \%,
an extremely low value for such a violent interaction.

\section{Discussion and conclusion}

These observations have major consequences in constraining the evolution of cluster galaxies. A high-velocity
off-center collision between two galaxies of relatively similar mass, whose violence is able to perturb the stellar 
distribution producing important tidal tails, is insufficient to induce a significant instantaneous starburst. This result might be representative
only of the nearby Universe where encounters of gas-rich galaxies are probably rare since clusters are dominated by gas-poor early-type galaxies 
such as the companion galaxy NGC 4435. It is conceivable, however, that at higher redshifts, where clusters are forming, stellar masses produced by a starburst
induced by interactions predicted by the models of Moore et al$.$ (1996; galaxy harassment) might be more important given
the higher fraction of gas-rich galaxies. \\
The other interesting result is the long time differential between the age of the interaction ($\sim$ 100 Myr as determined by dynamical
simulations, Combes et al$.$ 1988; Vollmer et al$.$ 2005) and the beginning of the starburst ($\sim$ 10 Myr in regions 1 and 7, $\sim$ 100 Myr 
in regions 2 and 5). This result is totally consistent with the models of Mihos et al$.$ (1991) that predict for close-by encounters 
an enhancement of the star formation activity in the inner disk during some 100 Myr, stopping once the gas reservoir is
exhausted as in NGC 4438.  
In the tidal tails, on the contrary, star formation is expected to increase after $\sim$ 100 Myr, the time needed 
by the gas to re-collapse, but then ceasing after
a few Myr because the expansion of the tidal tail brings the gas surface density to subcritical values 
(no HI and CO has been detected in these regions).
If these systems are dynamically stable and survive the interaction, they might be at the origin of some 
dwarf galaxies in the cluster similar to those observed in the Stephan's Quintet by Mendes de Oliveira et al$.$ (2004)
or in other interacting systems (Neff et al$.$ 2005; Hibbard et al$.$ 2005, Saviane et al$.$ 2004)
Being produced by a single starburst, these gas poor systems might evolve into dwarf ellipticals, typical of rich clusters. 
Otherwise they will simply increase the fraction of unbound stars, contributing to the Virgo 
intracluster light (Willman et al$.$ 2004).\\

\acknowledgements
GALEX (Galaxy Evolution Explorer) is a NASA Small Explorer, launched in April 2003.
We gratefully acknowledge NASA's support for construction, operation,
and science analysis for the GALEX mission,
developed in cooperation with the Centre National d'Etudes Spatiales
of France and the Korean Ministry of
Science and Technology. We wish to thank the referee, V. Charmandaris, for precious comments which helped
improving the quality of the manuscript.

\references

\reference{}Abazajian, K., et al$.$, 2004, astro-ph/0410239

\reference{}Boissier, S. \& Prantzos, N., 2000, MNRAS, 312, 398 

\reference{}Boissier, S., Prantzos, N., Boselli, A. \& Gavazzi, G., 2003, MNRAS, 346, 1215 

\reference{}Boselli, A. \& Gavazzi, G., 2002, A\&A, 386, 124

\reference{}Boselli, A., Tuffs, R., Gavazzi, G., Hippelein, H. \& Pierini, D., 1997, A\&AS, 121, 507

\reference{}Boselli, A., Gavazzi, G., Donas, J. \& Scodeggio, M., 2001, AJ, 121, 753

\reference{}Boselli, A., Gavazzi, G. \& Sanvito, G., 2003a, A\&A, 402, 37

\reference{}Boselli, A., Sauvage, M., Lequeux, J., Donati, A. \& Gavazzi, G., 2003b, A\&A, 406, 867

\reference{}Calzetti, D., 2001, PASP, 113, 1449

\reference{}Cayatte, V., van Gorkom, J., Balkowski, C., \& Kotanyi, C., 1990, AJ, 100, 604


\reference{}Combes, F., Dupraz, C., Casoli, F. \& Pagani, L., 1988, A\&A, 203, L9

\reference{}Draine, B.T., 2003, ApJ, 598, 1017

\reference{}Gavazzi, G., Boselli, A., Donati, A., Franzetti, P. \& Scodeggio, M., 2003, A\&A, 400, 451

\reference{}Gavazzi, G., Zaccardo, A., Sanvito, G., Boselli, A. \& Bonfanti, C., 2004, A\&A, 417, 499

\reference{}Jarrett, T., Chester, T., Cutri, R., Schneider, S. \& Huchra, J., 2003, AJ, 125, 525

\reference{}Hibbard, J., et al$.$, 2005, ApJL, Galex special issue (astro-ph/0411352)

\reference{}Hoopes, C., et al$.$, 2005, ApJL, Galex special issue (astro-ph/0411309)

\reference{}Hummel, E. \& Saikia, D., 1991, A\&A, 249, 43

\reference{}Keel, W. \& Wehrle, A., 1993, AJ, 106, 236

\reference{}Kenney, J. \& Yale, E., 2002, ApJ, 567, 865

\reference{}Kenney, J., Rubin, V., Planesas, P. \& Young, J., 1995, ApJ, 438, 135

\reference{}Kennicutt, R., 1998, ARA\&A, 36, 189

\reference{}Kong, X., Charlot, S., Brinchmann, J. \& Fall, S., 2004, MNRAS, 349, 769

\reference{}Kotanyi, C., van Gorkom, J. \& Ekers, R., 1983, ApJ, 273, L7

\reference{}Leitherer, C., et  al., 1999, ApJS, 123, 3 

\reference{}Machacek, M., Jones, C. \& Forman, W., 2004, ApJ, 610, 183

\reference{}Martin, C., et al$.$, 2005, ApJL, Galex special issue (astro-ph/0411302)

\reference{}Mendes de Oliveira, C., Cypriano, E., Sodr\'e, L., Balkowski, C., 2004, ApJ, 605, L20

\reference{}Mihos, C., Richstone, D., Bothun, G., 1991, ApJ, 377, 72

\reference{}Moore, B., Katz, N., Lake, G., Dressler, A. \& Oemler, A., 1996, Nat, 379, 613

\reference{}Morrissey, P., et al$.$, 2005, ApJL, Galex special issue (astro-ph/0411310)

\reference{}Neff, S., et al$.$, 2005, ApJL, Galex special issue (astro-ph/0411372)

\reference{}Saviane, I., Hibbard, J., Rich, M., 2004, AJ, 127, 660

\reference{}Vollmer, B., Braine, J., Combes, F. \& Sofue, Y., 2005, submitted to A\&A

\reference{}Willman, B., Governato, F., Wadsley, J. \& Quinn, 2004, MNRAS, 355, 159


\clearpage

\begin{figure*}
\epsscale{0.92}
\plotone{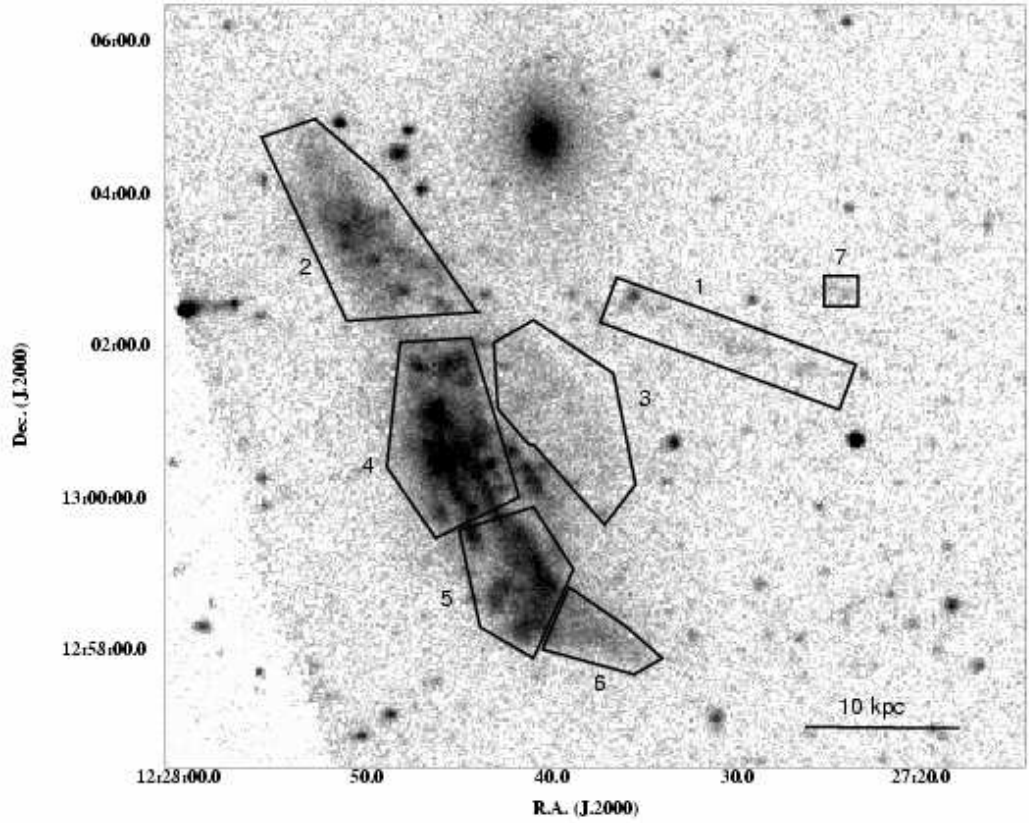}
\small{\caption{The combined NUV and FUV image of NGC 4438. The regions described in sect. 3 of the text are labeled 1 to 7.
The horizontal line is 10 kpc long (assuming a distance of 17 Mpc).}
\label{region}}
\end{figure*}

\begin{figure*}
\epsscale{0.92}
\plotone{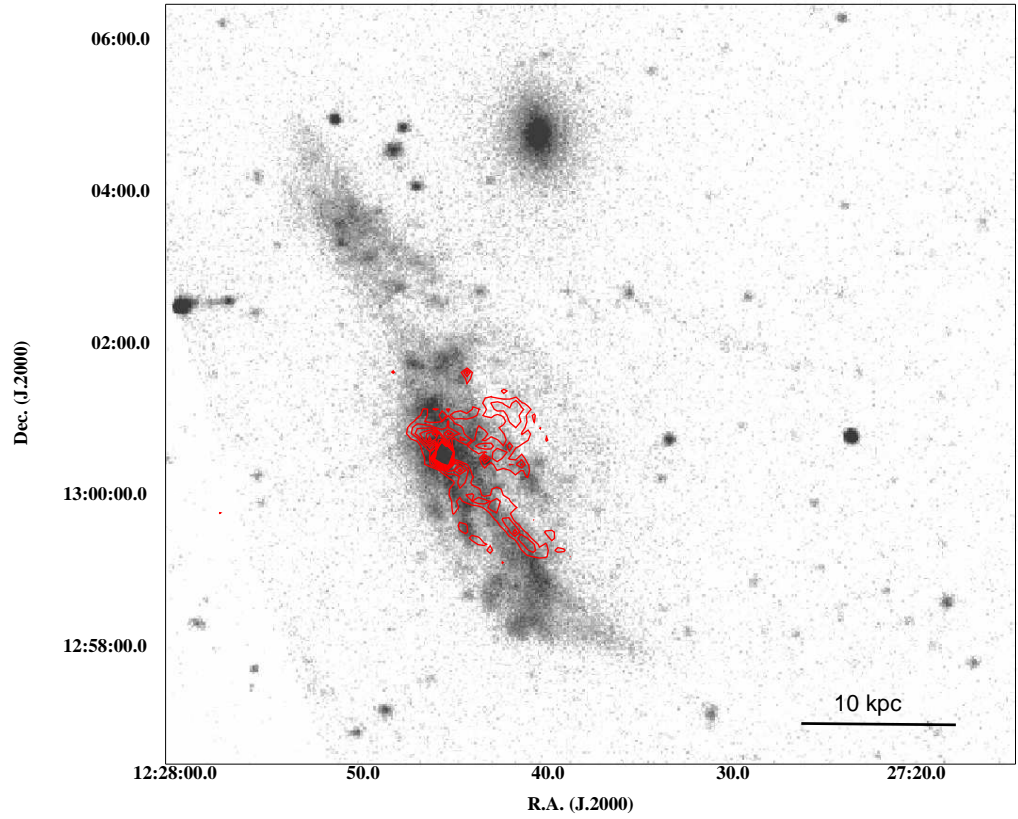}
\small{\caption{The H$\alpha$+[NII] contours (red, in arbitrary scale, in between 8 10$^{-17}$ and 6 10$^{-16}$  
erg cm$^{-2}$ s$^{-2}$ arcsec$^{-2}$, with $\sigma$= 5 10$^{-17}$  
erg cm$^{-2}$ s$^{-2}$ arcsec$^{-2}$, from Boselli \& Gavazzi (2002)) are superposed to the NUV gray-level image of NGC 4438.}
\label{contour}}
\end{figure*}

\begin{figure*}
\epsscale{1.25}
\plotone{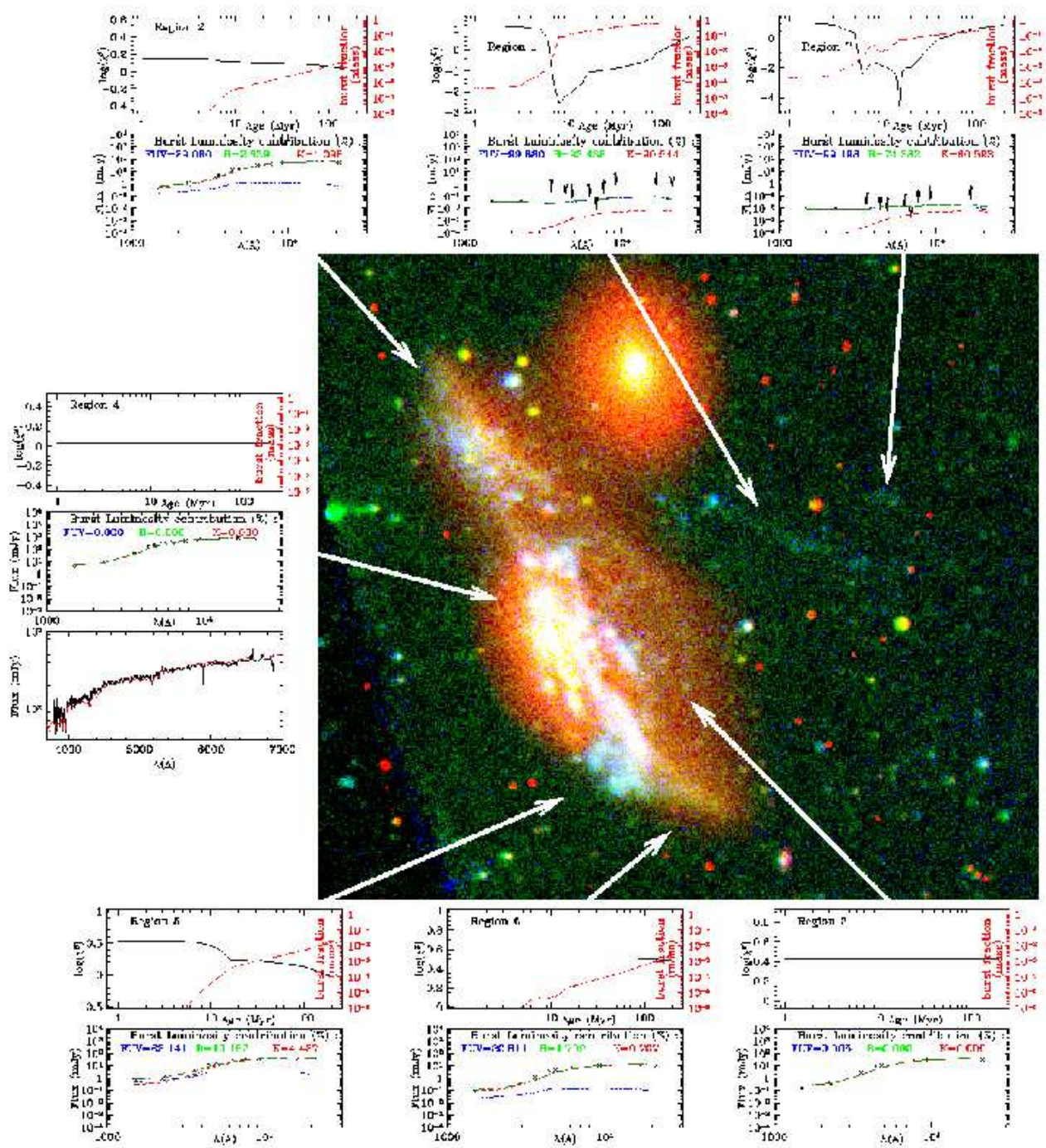}
\small{\caption{The RGB (FUV=blue, NUV=green, B=red) color map of NGC 4438 and NGC 4435. The SED of each 
region defined in Fig. 1 are given in the lower plot of each frame. Crosses indicate the observed data, arrows upper limits (in mJy), 
the red dashed line
the evolved population fit as determined by the model of Boissier \& Prantzos (2000), the dotted blue line the starburst SED (from Starburst
99) and the dashed green line the combined fitting model. The burst luminosity contribution (for the age corresponding to the
minimum $\chi$$^2$) in the band FUV, B and K is also given.
The upper panel gives the variation of the reduced $\chi$$^2$ parameter (black continuum line, in logarithmic scale) and of the burst mass fraction (red
dotted line) as a function of the age of the burst (in Myr). The lower panel of region 4 gives the integrated 3500 to 7000 \AA, $R$=1000
spectrum of the main body of the galaxy (black continuum line) compared to the fitted model (red dashed line).}
\label{RGB}}
\end{figure*}

\end{document}